\documentclass[aps,prd,preprintnumbers]{revtex4-2}
\usepackage{graphicx,float,wrapfig,subfigure}
\usepackage{amsfonts,amsmath,amssymb,amstext}
\usepackage{latexsym}
\usepackage{bm}
\usepackage{color}
\usepackage{comment}
\usepackage[normalem]{ulem}

\newcommand{\be}{\begin{equation}}
\newcommand{\ee}{\end{equation}}
\newcommand{\ba}{\begin{eqnarray}}
\newcommand{\ea}{\end{eqnarray}}

\definecolor{red}{rgb}{0.7,0,0}
\definecolor{green}{rgb}{0,0.5,0}

\usepackage[colorlinks=true,
citecolor=magenta,
linkcolor=black,
urlcolor=magenta]{hyperref}

\begin{document}

\title{RG-Consistent (P)NJL Model: Impact of Thermal Cutoff Modifications on Thermodynamics and Net-Baryon Number Fluctuations}
\date{\today}
\author{Jie Tang}
\affiliation{Center for Fundamental Physics, School of Mechanics and Physics, Anhui University of Science and Technology, Huainan, Anhui 232001, People's Republic of China}
\author{Fan Lin}
\email{linfan@aust.edu.cn (Corresponding Author)}
\affiliation{Center for Fundamental Physics, School of Mechanics and Physics, Anhui University of Science and Technology, Huainan, Anhui 232001, People's Republic of China}
\author{Xinyang Wang}
\email{wangxy@aust.edu.cn}
\affiliation{Center for Fundamental Physics, School of Mechanics and Physics, Anhui University of Science and Technology, Huainan, Anhui 232001, People's Republic of China}

\begin{abstract}
In this paper, we investigate the impact of renormalization group (RG) consistency on the chiral phase transition and thermodynamic properties of QCD matter using the RGNJL and RGPNJL models. By implementing a temperature-dependent thermal cutoff $\Lambda_T = k\Lambda_0$, we ensure that thermodynamic quantities converge toward the Stefan-Boltzmann limit at high temperatures, effectively extending the applicability of these effective theories. Our analysis shows that while the RG-consistency condition ($k \rightarrow \infty$) resolves causality violations in the RGNJL model by binding the speed of sound to the conformal limit, the RGPNJL model exhibits a more complex, non-monotonic sensitivity to the parameter $k$. Furthermore, we demonstrate that the RG-improved PNJL framework significantly enhances the description of net-baryon number fluctuations ($\kappa\sigma^2$) relative to lattice QCD data at vanishing chemical potential, though the intensification of these fluctuations at high baryon density highlights a critical sensitivity to the model's parametric constraints. This study provides a rigorous evaluation of the RG-consistency framework's predictive power in mapping the QCD phase diagram and interpreting experimental observables.

\end{abstract}

\maketitle

\section{Introduction}
\label{sec:1}
 \par The properties of Quantum Chromodynamics (QCD) matter at finite temperature and baryon density constitute one of the central frontiers in high-energy physics. Theoretical studies predict that QCD matter undergoes phase transitions as the temperature and baryon density increase~\cite{Fukushima:2010bq,McLerran:2007qj,Bazavov:2011nk}. Current theoretical understanding suggests that the chiral phase transition of QCD proceeds as a smooth crossover at low baryon chemical potential ($\mu_B$), while it is expected to turn into a first-order phase transition at larger $\mu_B$. The endpoint separating these two regimes is commonly referred to as the QCD critical point~\cite{Stephanov:1998dy,Fodor:2004nz}. Near this critical point, long-range correlations develop and the correlation length of the system grows significantly~\cite{Stephanov:1999zu}. As a consequence, event-by-event fluctuations of conserved charges become strongly enhanced~~\cite{Friman:2011pf}. In particular, the cumulants of the net-proton number, which are sensitive to the correlation length, are expected to exhibit characteristic critical behavior and may even diverge in the vicinity of the critical point. Over the past decade, extensive experimental efforts have therefore been devoted to searching for signatures of the QCD critical point through measurements of fluctuation observables~\cite{STAR:2013gus,STAR:2010vob}. These fluctuation observables, or equivalently the corresponding susceptibilities, provide an important experimental probe for mapping the structure of the QCD phase diagram. Recent experimental results~\cite{STAR:2020tga,STAR:2025zdq} report significant signals in the ratio $C_4/C_2$, which may serve as a characteristic indicator of nontrivial structures in the QCD phase diagram.

 \par In the theoretical regime, the infrared behavior of QCD is difficult to address due to its intrinsically non-perturbative nature. Lattice QCD provides a powerful first-principles approach; however, it still suffers from the well-known sign problem at finite baryon density. Consequently, effective field theories remain one of the most practical and reliable approaches for exploring non-perturbative aspects of QCD. Among these approaches, the Nambu--Jona--Lasinio (NJL) model~\cite{Nambu:1961tp,Nambu:1961fr,Klevansky:1992qe,Buballa:2003qv} is one of the most widely used low-energy effective theories of QCD, particularly for studying spontaneous chiral symmetry breaking and its restoration at finite temperature and density~\cite{Mao:2024rxh,Mei:2024rjg,Huang:2023ogw}. To incorporate aspects of gluon dynamics, the Polyakov loop is introduced, leading to the so-called Polyakov--Nambu--Jona--Lasinio (PNJL) model \cite{Fukushima:2003fw}. This extension enables a simultaneous investigation of the chiral phase transition and the confinement/deconfinement transition within an effective field theory framework \cite{Fukushima:2008wg,Fukushima:2010bq,Ratti:2006gh}. Nevertheless, as four-fermion interaction theories, both the NJL and PNJL models are non-renormalizable. As a consequence, the choice of regularization scheme becomes an essential component of the model and remains under active discussion. In practice, various cutoff schemes are employed to regularize the vacuum contribution, including hard or smooth cutoffs in three-momentum or four-momentum space~\cite{Klevansky:1992qe,Buballa:2003qv}. For the thermal contribution, however, it is still debated whether regularization is necessary; a detailed discussion can be found in Ref.~\cite{Yu:2015hym,Xue:2021ldz,Avancini:2019wed,Avancini:2018svs}.
 
 \par Recently, renormalization group (RG) consistency has been implemented in low-energy effective theories (LEFTs) to extend their applicability in hot and dense systems~\cite{Wetterich:1992yh,Braun:2018svj}. In general, a LEFT in vacuum is defined with a finite ultraviolet cutoff $\Lambda_{0}$; for the NJL model, one typically has $\Lambda_{0} \simeq 600\,\mathrm{MeV}$. However, in a medium with temperature or chemical potential exceeding $\Lambda_{0}$, contributions from higher momentum scales can become relevant, thereby challenging the validity of a fixed cutoff treatment. To maintain RG consistency, the full quantum effective action $\Gamma$ is required to be independent of the cutoff scale,
 \begin{equation}
 	\lim_{\Lambda \to +\infty} \Lambda \frac{d \Gamma}{d \Lambda} = 0 \,,
    \label{RGconsistency}
 \end{equation}
 which imposes nontrivial constraints on the ultraviolet regulator and ensures that physical observables remain insensitive to its specific choice. This framework has recently been applied to the NJL model (RG-improved NJL, or RGNJL) in Ref.~\cite{Gholami:2024diy,Gholami:2024ety} to investigate color superconductivity. Although the RG procedure extends the applicability of the non-renormalizable NJL model to hot and dense regimes, it remains unclear whether the RGNJL framework resolves the inconsistencies associated with cutoff artifacts identified in previous studies~\cite{Xue:2021ldz}. Furthermore, it is important to systematically examine how the RG implementation modifies the predictions of the NJL/PNJL models and to assess their consistency with the latest experimental observations.

 \par In this study, we present a systematic analysis of the RG-improved NJL (RGNJL) model, focusing on the impact of RG implementation on the chiral phase transition, thermodynamic properties, and baryon number susceptibilities. To facilitate a more direct comparison with experimental data and to incorporate confinement-related effects, we further extend the analysis to the RG-consistent Polyakov-loop-extended Nambu--Jona--Lasinio (RGPNJL) model. By confronting our results with recent experimental measurements, we assess the reliability and predictive power of the RG-consistency framework.
  
 \par The paper is organized as follows. In Sec.~\ref{sec:2}, we review the theoretical framework of the RG(P)NJL model and discuss the associated regularization schemes. In Sec.~\ref{sec:3}, we analyze the chiral phase transition within this framework. In Sec.~\ref{sec:4}, we present results for thermodynamic quantities and baryon number fluctuations. Finally, Sec.~\ref{sec:5} contains a summary and an outlook for future research.

\section{FORMALISM}
\label{sec:2}
\par In this section, we introduce the two-flavor RG-improved NJL/PNJL models, which are among the most widely used low-energy effective theories of QCD.
 
 \par The Lagrangian density of the two-flavor NJL model is given by
 \begin{equation}
 	\mathcal{L}_{\text{NJL}}=\bar{\psi}\left(i \gamma^\mu \partial_\mu-\hat{m}\right) \psi+G_S\left[(\bar{\psi} \psi)^2+\left(\bar{\psi} i \gamma^5 \bm{\tau} \psi\right)^2\right],
 \end{equation}
 where $\psi=(u,d)^{\mathrm{T}}$ denotes the two-flavor quark field, and $\hat{m}=\mathrm{diag}(m_u,m_d)$ is the current quark mass matrix. For simplicity, we assume isospin symmetry, $m_u = m_d = m$. The Pauli matrices $\bm{\tau}=(\tau^1,\tau^2,\tau^3)$ act in flavor space. The coupling constant $G_S$ characterizes the interaction strength in the scalar and pseudoscalar channels.
 
 \par Within the mean-field approximation, quantum fluctuations around the condensates are neglected, and the four-fermion interactions can be linearized as
 \begin{equation}
 	(\bar{\psi}\psi)^2
 	=2\langle\bar{\psi}\psi\rangle\,\bar{\psi}\psi
 	-\langle\bar{\psi}\psi\rangle^2,
 	\qquad
 	(\bar{\psi}i\gamma^{5}\boldsymbol{\tau}\psi)^2
 	=2\langle\bar{\psi}i\gamma^{5}\boldsymbol{\tau}\psi\rangle
 	\,\bar{\psi}i\gamma^{5}\boldsymbol{\tau}\psi
 	-\langle\bar{\psi}i\gamma^{5}\boldsymbol{\tau}\psi\rangle^2 .
 \end{equation}
 Here, $\langle\bar{\psi}\psi\rangle$ and $\langle\bar{\psi}i\gamma^{5}\boldsymbol{\tau}\psi\rangle$ denote the chiral and pseudoscalar condensates, respectively. In the absence of an axial chemical potential $\mu_5$, the pseudoscalar condensate vanishes, $\langle\bar{\psi}i\gamma^{5}\boldsymbol{\tau}\psi\rangle=0$. The Lagrangian density then reduces to
 \begin{equation}
 		\mathcal{L}_{\text{NJL}}
 		=\bar{\psi}\left(
 		i\gamma^{\mu}\partial_{\mu}-\hat{m}+2G_S\langle\bar{\psi}\psi\rangle
 		\right)\psi
 		-G_S\langle\bar{\psi}\psi\rangle^2=\bar{\psi}\left(
 		i\gamma^{\mu}\partial_{\mu}-M
 		\right)\psi
 		-\frac{(M-m)^2}{4G_S},
 \end{equation}
 where we have introduced the constituent quark mass $M = m - 2G_S\langle\bar{\psi}\psi\rangle$.

\par The constituent quark mass is determined self-consistently from the thermodynamic potential once finite temperature and chemical potential effects are taken into account. Following standard procedures~\cite{Klevansky:1992qe,Buballa:2003qv,Lin:2022ied}, the grand thermodynamic potential at the one-loop level for finite temperature $T$ and chemical potential of quark $\mu$ can be written as
\begin{equation}
	\Omega_{\mathrm{NJL}}=\frac{(M-m)^2}{4G_S}
	-2 N_{c} N_{f} \int \frac{d^{3} p}{(2 \pi)^{3}}
	\left[
	E
	+T \ln \left(1+e^{-\beta(E-\mu)}\right)
	+T \ln \left(1+e^{-\beta(E+\mu)}\right)
	\right],
\end{equation}
where $\beta=1/T$ and $E=\sqrt{p^{2}+M^{2}}$ denotes the single-particle dispersion relation of the constituent quark. The physical value of the constituent quark mass is obtained by minimizing the thermodynamic potential, which is equivalent to solving the gap equation
\begin{equation}
	\frac{d \Omega_{\mathrm{NJL}}}{d M}=0,
\end{equation}
However, this equation cannot be solved without further specification, since the NJL model is non-renormalizable and the momentum integral is ultraviolet divergent. Therefore, an appropriate regularization scheme must be introduced before proceeding with the numerical evaluation.

\par In practice, the ultraviolet divergence in the vacuum contribution is typically treated by introducing a momentum cutoff. A common choice is the three-momentum regularization scheme. For instance, in the hard-cutoff scheme, the upper limit of the momentum integral is replaced by a finite cutoff $\Lambda$, i.e., the integration range is restricted from $[0,\infty)$ to $[0,\Lambda]$. In contrast, for the thermal contribution to the thermodynamic potential, it remains under debate whether a similar cutoff should be imposed. As discussed in Ref.~\cite{Xue:2021ldz}, applying the cutoff to the finite (thermal) part of the integral may lead to inconsistencies, and different prescriptions can yield qualitatively different results. Recently, the implementation of RG consistency in effective theories has provided a more systematic framework to address this issue and further clarify the role of the cutoff in both vacuum and thermal contributions.

\par As shown in Ref.~\cite{Gholami:2024diy}, imposing the RG consistency condition in Eq.~(\ref{RGconsistency}) on the NJL model modifies the regularization procedure. In particular, an additional cutoff scale is introduced for the thermal contribution, which can be parameterized as $\Lambda_{T}=k\,\Lambda_{0}$. In general, for a medium with temperature or chemical potential exceeding the vacuum cutoff scale $\Lambda_{0}$, contributions from higher momentum modes become relevant. This implies that the parameter $k$ should satisfy $k \geq 1$, with its precise value determined by the physical conditions under consideration. Accordingly, the grand thermodynamic potential of the RG-improved NJL (RGNJL) model can be expressed as
\begin{equation}
	\Omega _ {\mathrm{RGNJL}} = \frac{(M-m)^2}{4G_S}
	-2N_{c}N_{f}\int_{0}^{\Lambda_{0}}\frac{d^{3}p}{(2\pi)^{3}}E
	-2N_{c}N_{f}T\int_{0}^{\Lambda_{T}}\frac{d^{3}p}{(2\pi)^{3}}
	\left[
	\ln\left(1+e^{-\beta(E-\mu)}\right)
	+\ln\left(1+e^{-\beta(E+\mu)}\right)
	\right],
\end{equation}
where $\Lambda_{T}=k\,\Lambda_{0}$ explicitly separates the cutoff scales for the vacuum and thermal contributions.

\par While the NJL model successfully describes the chiral phase transition, it lacks confinement dynamics. This limitation motivates the extension to the RG-consistent two-flavor PNJL model. The order parameter for the confinement/deconfinement phase transition is the Polyakov loop,
\begin{equation}
	L(\vec{x})=\mathcal{P}\exp\left[i\int_0^{1/T}d\tau\, A_4(\vec{x},\tau)\right],
\end{equation}
where $A_4$ denotes the temporal component of the background gluon field. The traced Polyakov loop and its conjugate are defined as
\begin{equation}
	\Phi=\frac{\mathrm{Tr}_c\,L}{N_c},\quad \bar{\Phi}=\frac{\mathrm{Tr}_c\,L^\dagger}{N_c},
\end{equation}
which effectively act as imaginary chemical potentials for quarks and can be incorporated into the thermodynamic potential~\cite{Weiss:1980rj,Weiss:1981ev,Fukushima:2000ww,Fukushima:2003fw}.

\par The grand thermodynamic potential of the RGPNJL model then takes the form
\begin{equation}
	\begin{aligned}
		\Omega _ {\mathrm{RGPNJL}} &= \frac{(M-m)^2}{4G_S}+\mathcal{U}(\Phi,\bar{\Phi},T)
		-2N_{c}N_{f}\int_{0}^{\Lambda_{0}}\frac{d^{3}p}{(2\pi)^{3}}E \\
		&-2N_fT\int_{0}^{\Lambda_T}\frac{d^3p}{(2\pi)^3}\ln\left(1+3\Phi e^{-\beta(E-\mu)}+3\bar{\Phi}e^{-2\beta(E-\mu)}+e^{-3\beta(E-\mu)}\right)\\
		&-2N_fT\int_{0}^{\Lambda_T}\frac{d^3p}{(2\pi)^3}\ln\left(1+3\bar{\Phi} e^{-\beta(E+\mu)}+3\Phi e^{-2\beta(E+\mu)}+e^{-3\beta(E+\mu)}\right),
	\end{aligned}
\end{equation}
where the same separation of cutoff scales has been applied to the thermal contributions. The effective potential $\mathcal{U}(\Phi,\bar{\Phi},T)$ encodes the gluonic degrees of freedom and cannot be derived directly from QCD. A commonly used phenomenological parametrization is~\cite{Weiss:1980rj,Ratti:2006wg,Shao:2017yzv}
\begin{equation}
	\frac{\mathcal{U}(\Phi,\bar{\Phi},T)}{T^4}
	=-\frac{a(T)}{2}\bar{\Phi}\Phi
	+b(T)\ln\left[1-6\bar{\Phi}\Phi +4(\bar{\Phi}^3+\Phi^3)-3(\bar{\Phi}\Phi)^2\right],
\end{equation}
where $a(T)=a_0+a_1\left(\frac{T_0}{T}\right)+a_2\left(\frac{T_0}{T}\right)^2$ and $b(T)=b_3\left(\frac{T_0}{T}\right)^3$.

\par Before solving the RGNJL/RGPNJL models and analyzing the corresponding physical effects, the model parameters must be fixed to reproduce basic hadronic observables in vacuum, such as the pion mass $m_\pi=139.3\,\mathrm{MeV}$ and the pion decay constant $f_\pi=92\,\mathrm{MeV}$. In this work, we adopt the standard parameter set $m_u = m_d = 5.5\,\mathrm{MeV},  N_c = 3,  N_f = 2 $
together with $\Lambda_{0}=651\,\mathrm{MeV}, G_S=5.04\,\mathrm{GeV}^{-2} $ which are used consistently in both the RGNJL and RGPNJL models. For the RGPNJL model, the parameters in the Polyakov-loop effective potential are chosen as $a_0=3.51, a_1=-2.47, a_2=15.22, b_3=-1.75$ which are fixed phenomenologically to reproduce lattice QCD results in the pure gauge sector~\cite{Weiss:1980rj,Ratti:2006wg,Shao:2017yzv}.

\section{Chiral phase transition}
\label{sec:3}

\par The physical constituent quark mass is determined by minimizing the thermodynamic potential. By solving the corresponding gap equation, one obtains the constituent quark mass as a function of temperature and chemical potential for a given RG-modified thermal cutoff $\Lambda_T = k\,\Lambda_{0}$.

\par For the RGNJL model, the gap equation reads
\begin{equation}
	\frac{\partial\Omega_{\text{RGNJL}}}{\partial M}
	= \frac{M-m}{2G_S}
	-2N_{c}N_{f}\int_{0}^{\Lambda_{0}}\frac{d^{3}p}{(2\pi)^{3}}\frac{M}{E}
	+2N_{c}N_{f}\int_{0}^{\Lambda_{T}}\frac{d^{3}p}{(2\pi)^{3}}
	\frac{M}{E}\left[
	\frac{1}{1+e^{\beta(E-\mu)}}+\frac{1}{1+e^{\beta(E+\mu)}}
	\right]
	= 0.
\end{equation}

\par Similarly, for the RGPNJL model, the stationarity conditions with respect to $M$, $\Phi$, and $\bar{\Phi}$ lead to a coupled set of gap equations. The gap equation for the constituent quark mass is given by
\begin{equation}
	\begin{aligned}
		\frac{\partial\Omega_{\text{RGPNJL}}}{\partial M}
		=&\frac{M-m}{2G_S}
		-2 N_c N_{f}\int_{0}^{\Lambda_0}\frac{d^{3}p}{(2\pi)^{3}}\frac{M}{E}
		\\
		&+2N_{f}\int_{0}^{\Lambda_{T}}\frac{d^{3}p}{(2\pi)^{3}}\frac{M}{E}
		\frac{3\Phi e^{-\beta(E-\mu)}+6\bar{\Phi}e^{-2\beta(E-\mu)}+3e^{-3\beta(E-\mu)}}
		{1+3\Phi e^{-\beta(E-\mu)}+3\bar{\Phi}e^{-2\beta(E-\mu)}+e^{-3\beta(E-\mu)}}
		\\
		&+2N_{f}\int_{0}^{\Lambda_{T}}\frac{d^{3}p}{(2\pi)^{3}}\frac{M}{E}
		\frac{3\bar{\Phi} e^{-\beta(E+\mu)}+6\Phi e^{-2\beta(E+\mu)}+3e^{-3\beta(E+\mu)}}
		{1+3\bar{\Phi} e^{-\beta(E+\mu)}+3\Phi e^{-2\beta(E+\mu)}+e^{-3\beta(E+\mu)}}
		=0 \,.
	\end{aligned}
\end{equation}

\par The stationarity conditions with respect to the Polyakov loop variables $\Phi$ and $\bar{\Phi}$ are given by
\begin{equation}
	\begin{aligned}
		\frac{\partial\Omega_{\text{RGPNJL}}}{\partial \Phi }
		=&T^4\left[
		-\frac{a(T)}{2}\bar{\Phi}
		+b(T)\frac{-6\bar{\Phi}+12\Phi^2-6\bar{\Phi}^2\Phi}
		{1-6\bar{\Phi}\Phi+4\left(\bar{\Phi}^3+\Phi^3\right)-3(\bar{\Phi}\Phi)^2}
		\right]
		\\
		&-2N_fT\int_{0}^{\Lambda_T}\frac{d^3p}{(2\pi)^3}
		\frac{3e^{-\beta(E-\mu)}}
		{1+3\Phi e^{-\beta(E-\mu)}+3\bar{\Phi}e^{-2\beta(E-\mu)}+e^{-3\beta(E-\mu)}}
		\\
		&-2N_{f}T\int_{0}^{\Lambda_T}\frac{d^{3}p}{(2\pi)^{3}}
		\frac{3e^{-\beta(E+\mu)}}
		{1+3\Phi e^{-\beta(E+\mu)}+3\bar{\Phi}e^{-2\beta(E+\mu)}+e^{-3\beta(E+\mu)}}
		=0 \,,
	\end{aligned}
\end{equation}

\begin{equation}
	\begin{aligned}
		\frac{\partial\Omega_{\text{RGPNJL}}}{\partial \bar{\Phi} }
		=&T^4\left[
		-\frac{a(T)}{2}\Phi
		+b(T)\frac{-6\Phi+12\bar{\Phi}^2-6\bar{\Phi}\Phi^2}
		{1-6\bar{\Phi}\Phi+4\left(\bar{\Phi}^3+\Phi^3\right)-3(\bar{\Phi}\Phi)^2}
		\right]
		\\
		&-2N_fT\int_{0}^{\Lambda_T}\frac{d^3p}{(2\pi)^3}
		\frac{3e^{-2\beta(E-\mu)}}
		{1+3\Phi e^{-\beta(E-\mu)}+3\bar{\Phi}e^{-2\beta(E-\mu)}+e^{-3\beta(E-\mu)}}
		\\
		&-2N_fT\int_{0}^{\Lambda_T}\frac{d^3p}{(2\pi)^3}
		\frac{3e^{-2\beta(E+\mu)}}
		{1+3\Phi e^{-\beta(E+\mu)}+3\bar{\Phi}e^{-2\beta(E+\mu)}+e^{-3\beta(E+\mu)}}
		=0 \,.
	\end{aligned}
\end{equation}

\par By solving the gap equations of the NJL/PNJL models numerically, one obtains the constituent quark mass as a function of temperature and chemical potential for different values of the thermal cutoff $\Lambda_T$. The corresponding results are shown in Fig. 1 and Fig. 2, for the NJL and PNJL models, respectively. The critical temperature of the chiral phase transition, $T_c$, can be defined through the condition
\begin{equation}
	T_c:\quad \left|\frac{dM}{dT}\right|_{T=T_c} = \max\left|\frac{dM}{dT}\right| \,,
\end{equation}
which characterizes the location of the most rapid change in the constituent quark mass.

\begin{figure}[!htbp]
	\centering
	\subfigure[]{
		\includegraphics[scale=0.45]{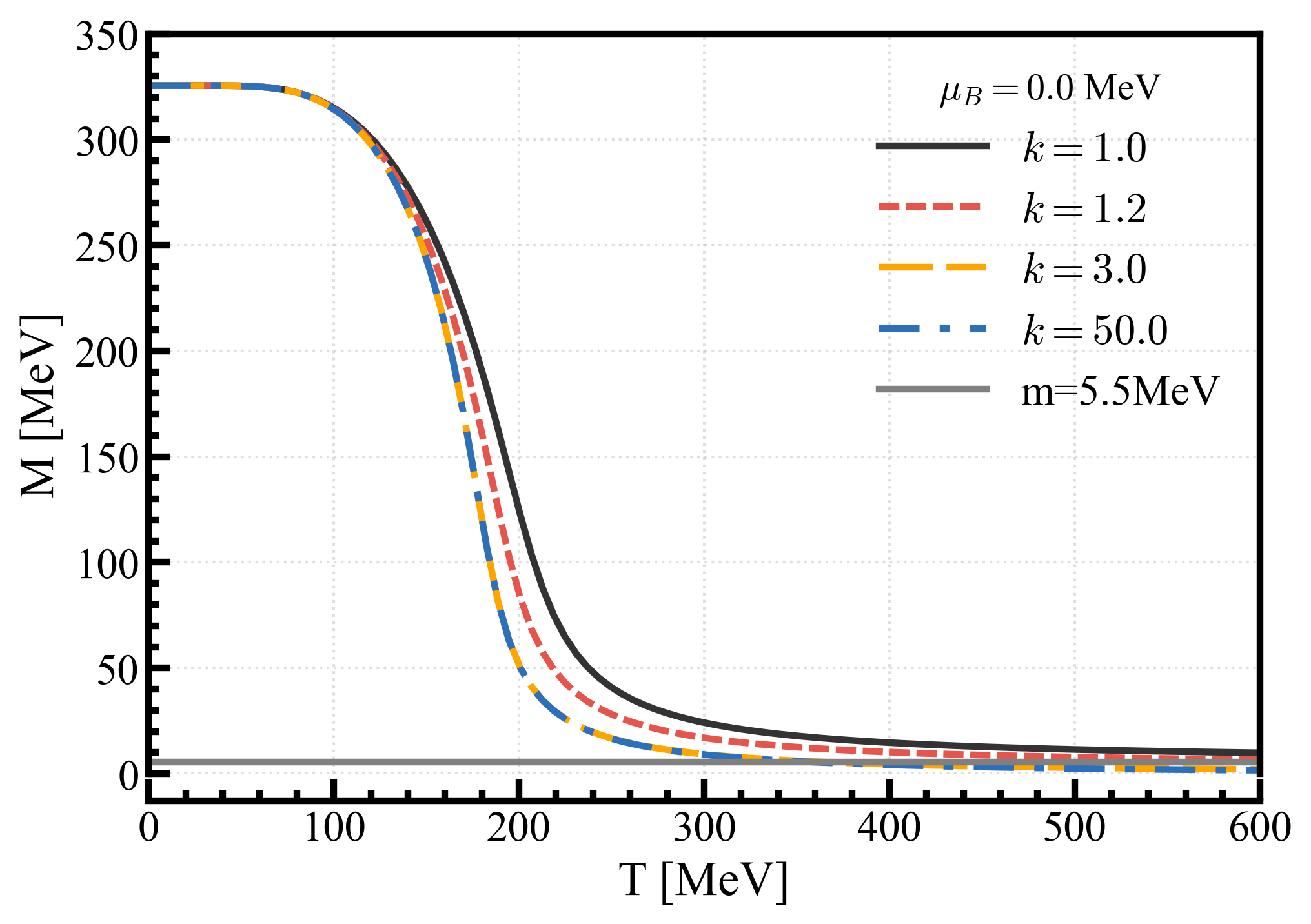}}
	\subfigure[]{
		\includegraphics[scale=0.45]{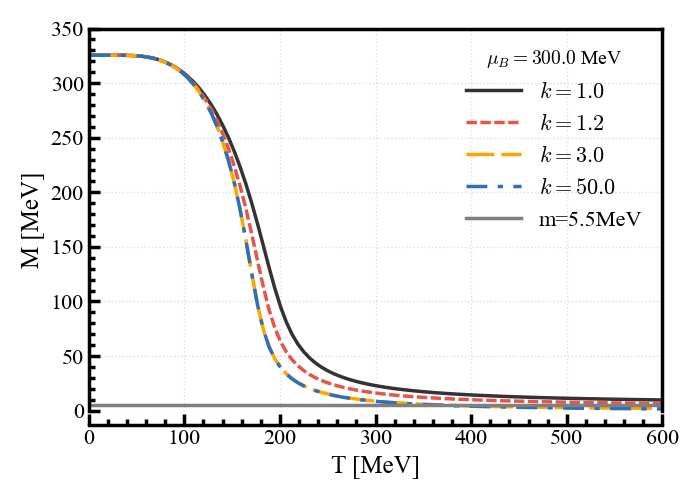}}
	\subfigure[]{
		\includegraphics[scale=0.45]{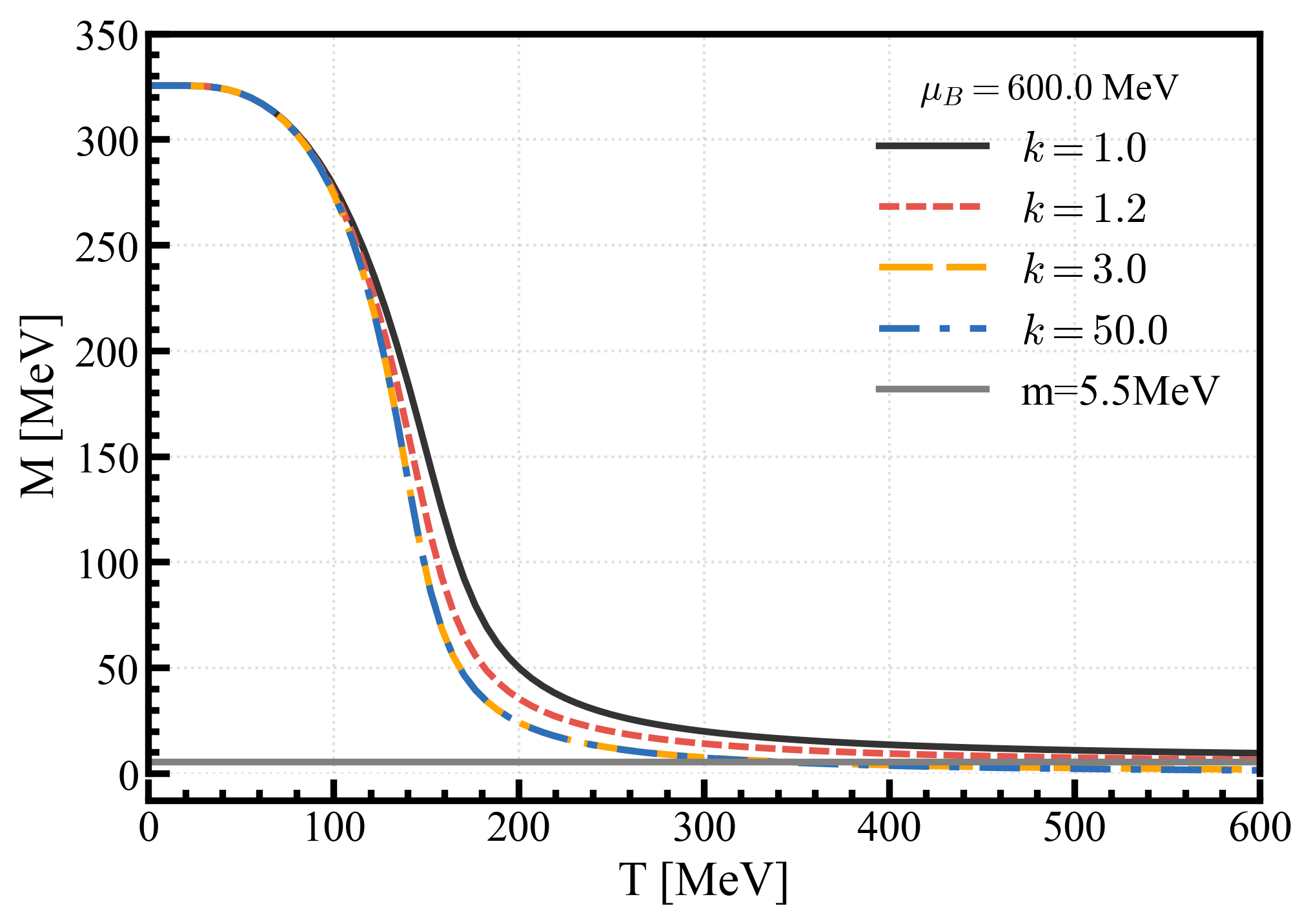}}
	\subfigure[]{
		\includegraphics[scale=0.45]{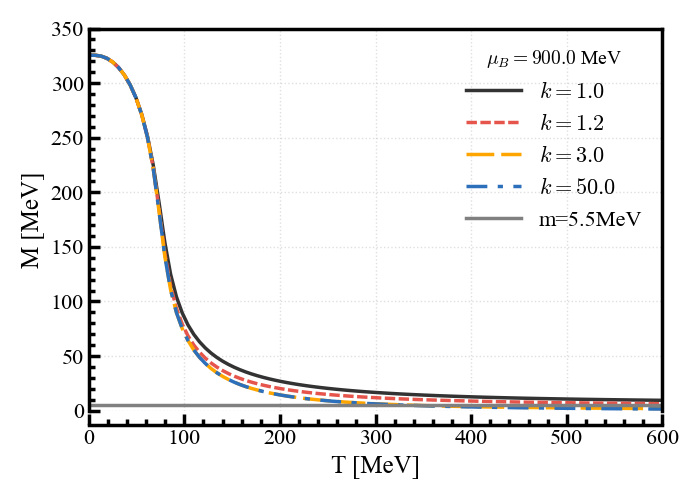}}
	\caption{Constituent quark mass $M$ as a function of temperature $T$ for different truncation factors $k$ (i.e., $\Lambda_T = k\Lambda_0$) at fixed baryon chemical potentials $\mu_B=0,\,300,\,600,$ and $900~\mathrm{MeV}$ in the RGNJL model.}
	\label{NJLTfixmu}
\end{figure}

\par
\begin{figure}[!htbp]
	\centering
	\subfigure[]{
		\includegraphics[scale=0.45]{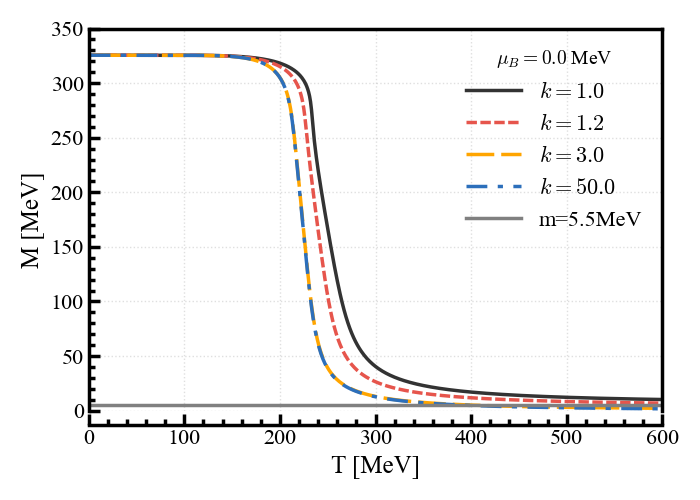}}
	\subfigure[]{
		\includegraphics[scale=0.45]{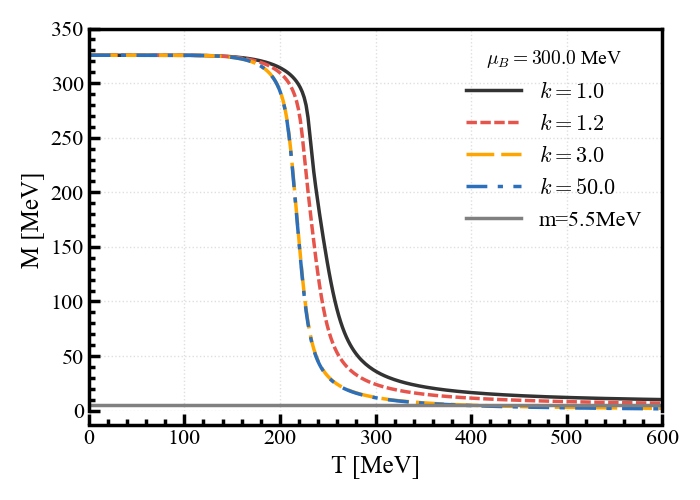}}
	\subfigure[]{
		\includegraphics[scale=0.45]{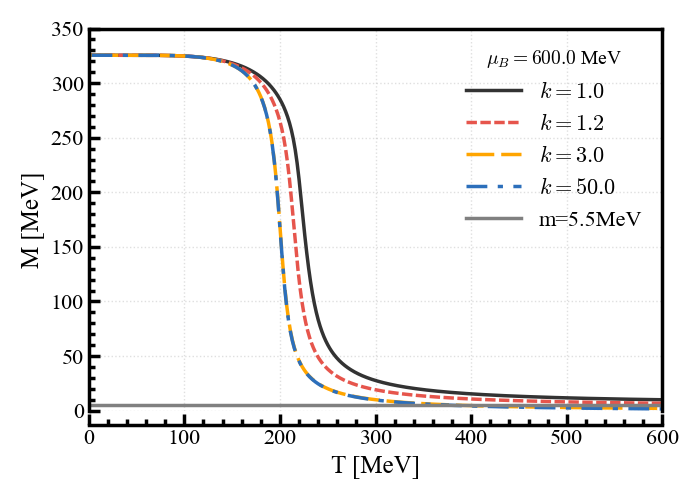}}
	\subfigure[]{
		\includegraphics[scale=0.45]{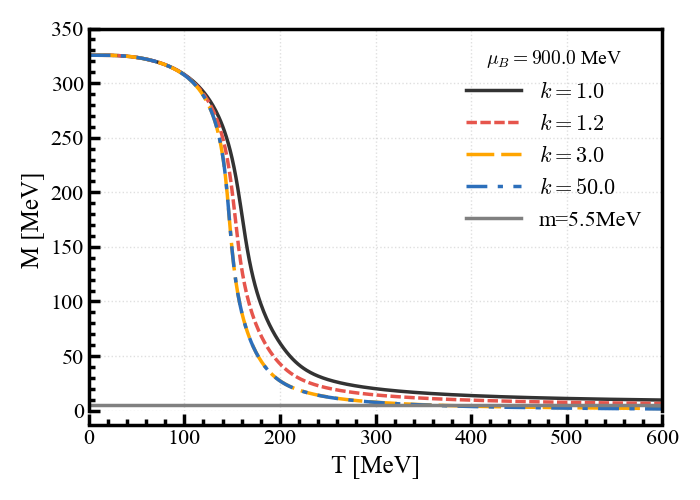}}
	\caption{Constituent quark mass $M$ as a function of temperature $T$ for different truncation factors $k$ (i.e., $\Lambda_T = k\Lambda_0$) at fixed baryon chemical potentials in the RGPNJL model, illustrating the influence of confinement effects.}
	\label{PNJLTfixmu}
\end{figure}

 As shown in Fig. \ref{NJLTfixmu} and Fig. \ref{PNJLTfixmu}, both the NJL and PNJL models exhibit a similar qualitative behaviorm. In the low-temperature regime, the constituent quark mass remains almost constant and shows only a weak dependence on the choice of $\Lambda_T$. This reflects the dominance of the vacuum contribution, where thermal effects are negligible. As the temperature approaches the transition region, however, the dependence on $\Lambda_T$ becomes pronounced. Larger values of $\Lambda_T$ lead to an earlier and more rapid decrease of the constituent quark mass, corresponding to a lower pseudo-critical temperature $T_c$. This behavior indicates that increasing the thermal cutoff effectively enhances the contribution of high-momentum thermal modes, thereby accelerating the restoration of chiral symmetry.

\par A more detailed inspection of the figures reveals an additional feature. For the case $k=1.0$, corresponding to $\Lambda_T=\Lambda_0$, the constituent quark mass decreases with increasing temperature but remains above the current quark mass $m$ even at high temperature. In contrast, for larger values of $k$, the constituent quark mass can drop below the current quark mass in the high-temperature regime. This behavior is unphysical, since the current quark mass sets the lower bound of the dynamical mass in the chirally restored phase. 

\par
\begin{figure}
	\centering
	\subfigure[]{
		\includegraphics[scale=0.45]{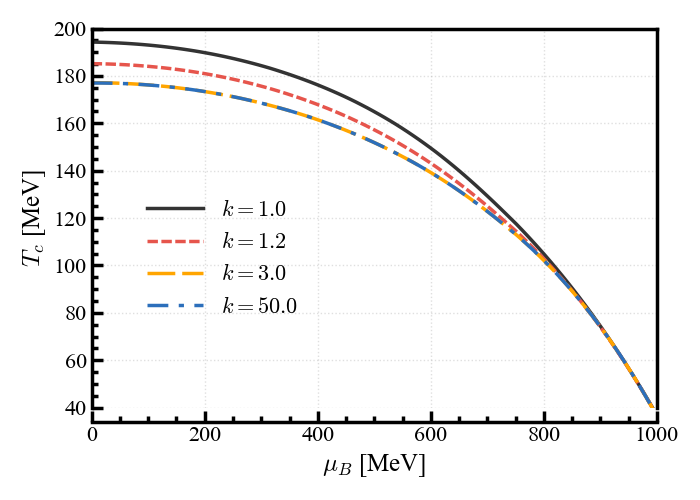}}
	\subfigure[]{
		\includegraphics[scale=0.45]{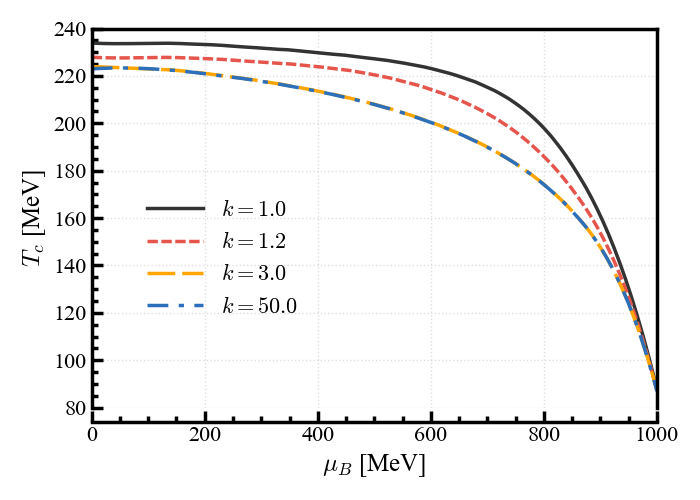}}
	\caption{Chiral phase transition critical temperature $T_c$ versus baryon chemical potential $\mu_B$ for the RGNJL (a) and RGPNJL (b) models at different $k$ values.}
	\label{Tc_muB}
\end{figure}

 Beyond the temperature-dependent mass evolution, the structure of the chiral phase boundary—characterized by the critical temperature $T_c$ as a function of the chemical potential of baryon $\mu_B=3 \mu $ —provides further insight into the framework. As illustrated in Fig.~\ref{Tc_muB}, both the RGNJL (a) and RGPNJL (b) models exhibit a monotonically decreasing phase boundary as the chemical potential increases. Notably, increasing the parameter $k$ results in a systematic suppression of $T_c$ across the entire range of $\mu_B$. However, this suppression is more pronounced at low chemical potentials; as $\mu_B$ approaches 1000 MeV, the curves for varying $k$ tend to converge, suggesting that the sensitivity of the chiral transition temperature to the RG-consistent modification diminishes at high baryon densities.

\par Therefore, although the RG-consistent extension modifies the thermal contributions and shifts the chiral transition, it does not fully eliminate the artifacts associated with the cutoff prescription. In particular, the appearance of $M < m$ at large $k$ suggests that the RGNJL/RGPNJL framework still requires further refinement to achieve a fully consistent description of the high-temperature regime.

\section{Thermodynamic quantities and baryon number fluctuations}
\label{sec:4}
\par In addition to the chiral phase transition and the behavior of the constituent quark mass, it is instructive to analyze thermodynamic observables that can be directly derived from the grand potential. These quantities provide complementary information on the equation of state and the response of the system to thermal excitations, and are therefore essential for a comprehensive understanding of the QCD medium~\cite{Ratti:2004ra,Zhang:2018ouu,Ghosh:2006qh}.

\par The pressure is defined as the negative of the grand potential density, $P=-\Omega$. To ensure a physically meaningful normalization, one subtracts the vacuum contribution such that the pressure vanishes at zero temperature and chemical potential. Accordingly, the pressure is given by
\begin{equation}
	P(\mu, T)=\Omega(0,0)-\Omega(\mu, T),
\end{equation}
where $\Omega(0,0)$ denotes the grand potential in vacuum.

\par The energy density $\epsilon$ can be obtained from standard thermodynamic relations,
\begin{equation}
	\epsilon=-\left.T^{2}\frac{\partial(\Omega/T)}{\partial T}\right|_{V}
	=-\left.T\frac{\partial\Omega}{\partial T}\right|_{V}+\Omega-\Omega(0,0),
\end{equation}
which explicitly incorporates the vacuum subtraction to maintain consistency with the normalization of the pressure.

\par The specific heat at constant volume, which characterizes the response of the system to temperature variations, is defined as
\begin{equation}
	C_{V}=\left.\frac{\partial\epsilon}{\partial T}\right|_{V}
	=-\left.T\frac{\partial^{2}\Omega}{\partial T^{2}}\right|_{V}.
\end{equation}
This quantity is particularly sensitive to the chiral phase transition and typically exhibits a peak structure near the pseudo-critical temperature.

\par Another important observable is the squared speed of sound, which encodes the stiffness of the equation of state and plays a crucial role in the hydrodynamic evolution of the medium. At constant entropy, it is given by
\begin{equation}
	v_{s}^{2} = \left. \frac{\partial P}{\partial \epsilon} \right|_{S}
	=  \left. \frac{\partial \Omega}{\partial T} \right|_{V} \Bigg/
	{ \left. T \frac{\partial^{2} \Omega}{\partial T^{2}} \right|_{V} }.
\end{equation}

\begin{figure}[!htbp]
\centering
 \subfigure[]{
 		\includegraphics[scale=0.45]{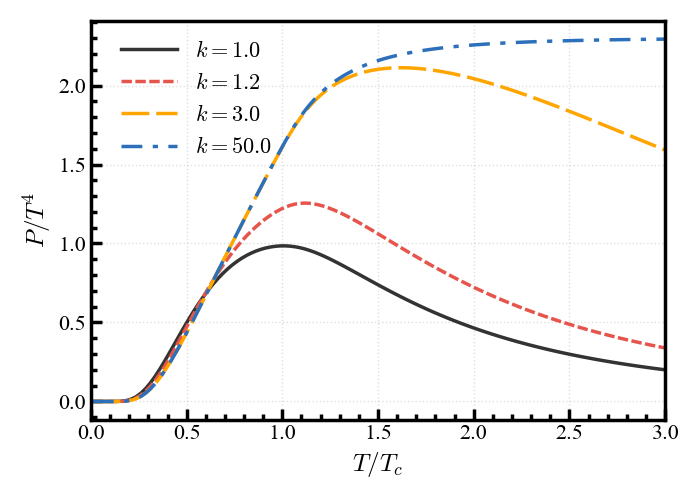}}
 \subfigure[]{
 		\includegraphics[scale=0.45]{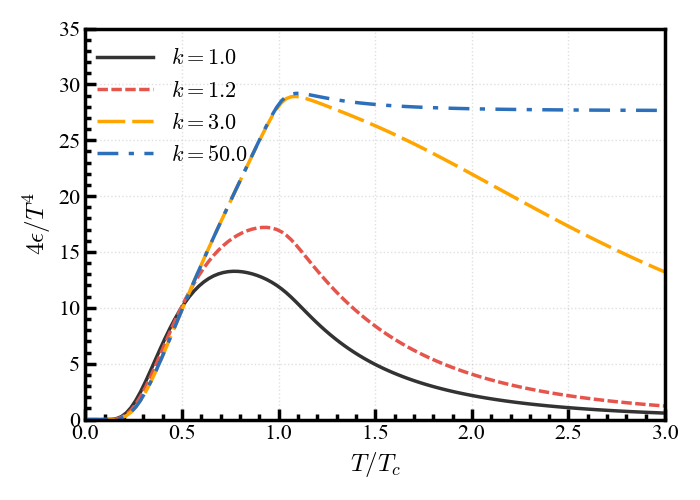}}
 \subfigure[]{
 		\includegraphics[scale=0.45]{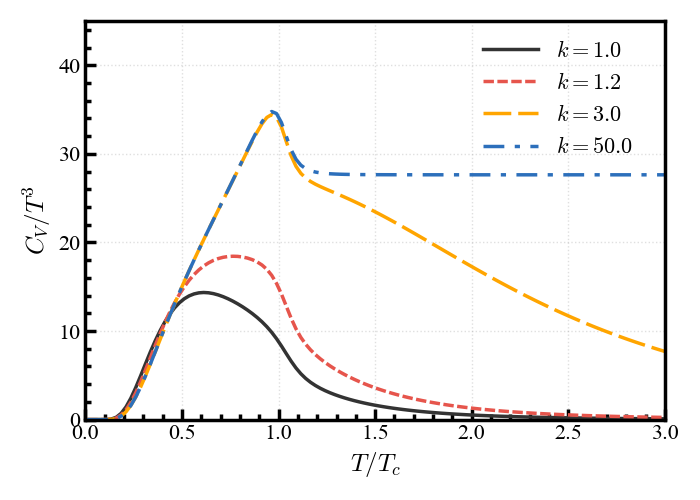}}
 \subfigure[]{
 		\includegraphics[scale=0.45]{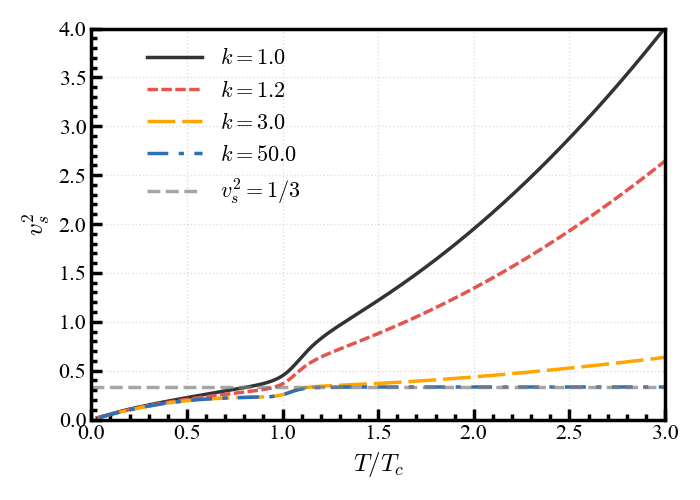}}
 	\caption{Thermodynamic quantities as a function of temperature at zero chemical potential for the RGNJL model.}
 	\label{NJL4}
 \end{figure}

\par
\begin{figure}[!htbp]
\centering
 \subfigure[]{
 		\includegraphics[scale=0.45]{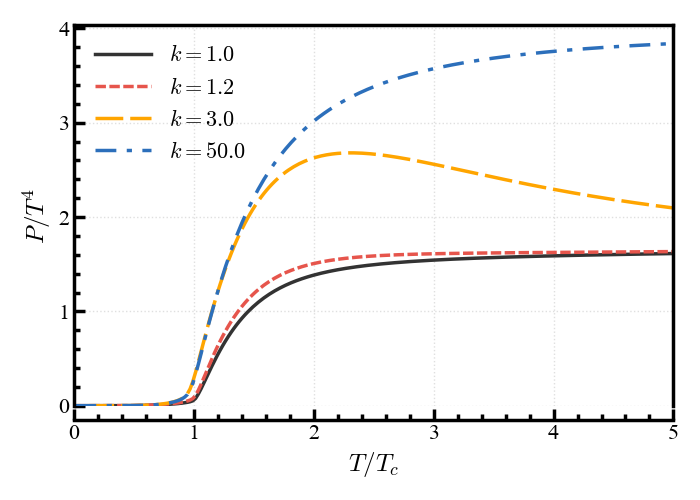}}
 \subfigure[]{
 		\includegraphics[scale=0.45]{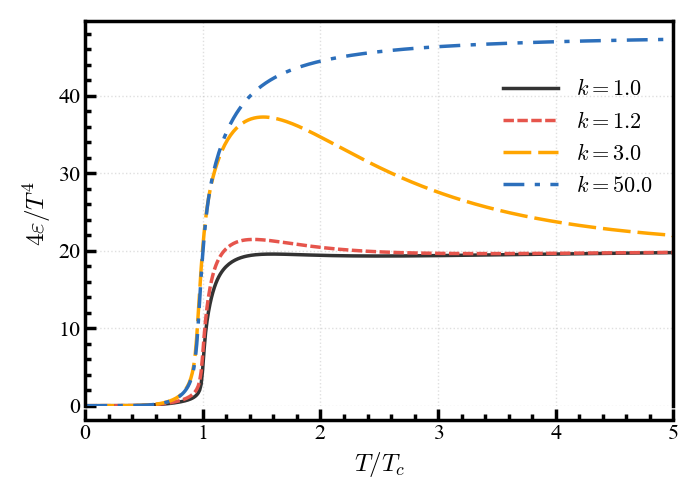}}
 \subfigure[]{
 		\includegraphics[scale=0.45]{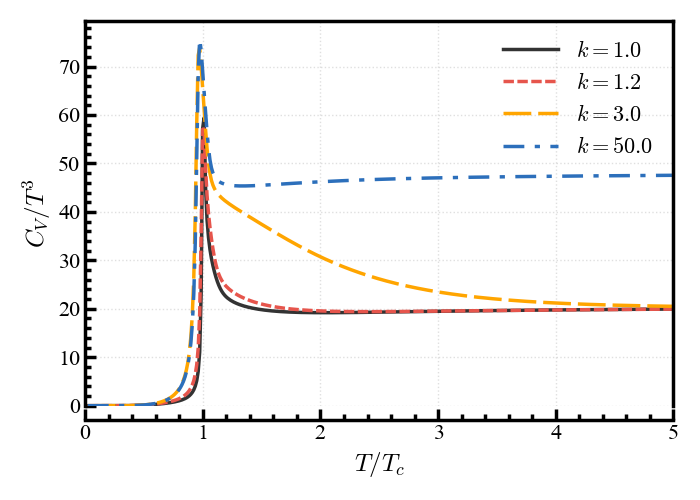}}
 \subfigure[]{
 		\includegraphics[scale=0.45]{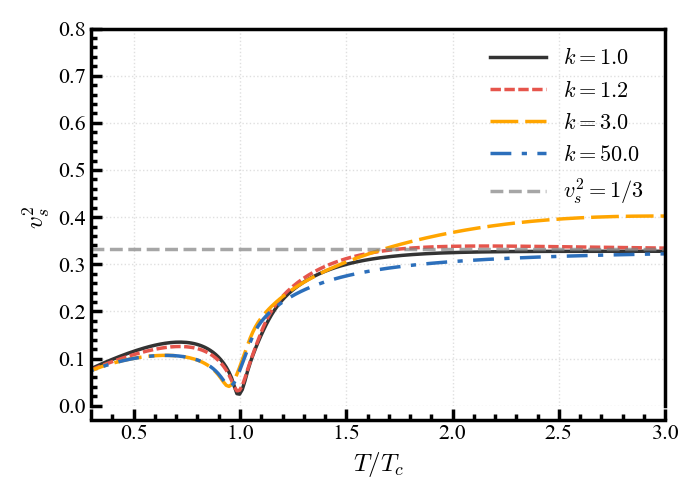}}
 	\caption{Thermodynamic quantities as a function of temperature at zero chemical potential for the RGPNJL model.}
 	\label{PNJL4}
 \end{figure}

 These thermodynamic quantities serve as independent probes to assess the impact of RG-consistent modifications on the equation of state and to evaluate the validity of the RGNJL and RGPNJL frameworks. As illustrated in Fig.~\ref{NJL4} and Fig.~\ref{PNJL4}, both models exhibit a systematic increase in the scaled pressure ($P/T^4$), energy density ($4\epsilon/T^4$), and specific heat ($C_V/T^3$) in the high-temperature regime as the parameter $k$ increases, eventually approaching a stable plateau. This behavior aligns with theoretical expectations: at high temperatures, the coupling becomes weak, and the thermal quantities converge toward the Stefan-Boltzmann limit of an ideal, free Fermi gas.

\par However, a comparison between the two models reveals distinct differences in the transition dynamics. In the RGNJL model (Fig.~\ref{NJL4}a-c), the thermodynamic quantities exhibit relatively smooth, gradual growth as $T/T_c$ increases. Conversely, the RGPNJL model (Fig.~\ref{PNJL4}a-c) displays significantly steeper growth rates, particularly in the range $0.8 \le T/T_c \le 1.2$. The specific heat $C_V/T^3$ in Fig.~\ref{PNJL4}c exhibits a very sharp, pronounced peak near the pseudo-critical temperature $T_c$, which is notably more aggressive than the broader, muted peak observed in the RGNJL model (Fig.~\ref{NJL4}c). This indicates that the inclusion of the Polyakov loop potential in the RGPNJL framework more effectively simulates the rapid change in degrees of freedom associated with the confinement-deconfinement transition.

\par The behavior of the squared speed of sound ($v_s^2$) presents a more complex scenario with distinct implications for causality and conformality. In the RGNJL model (Fig.~\ref{NJL4}d), the baseline ($k=1.0$) shows an unphysical violation of causality where $v_s^2 > 1$ at high temperatures. However, enforcing the RG consistency condition ($k \rightarrow \infty$) successfully binds $v_s^2$ to the conformal limit of $1/3$, resolving the violation. 

\par The RGPNJL model (Fig.~\ref{PNJL4}d) demonstrates a more intricate, non-monotonic dependence on $k$. While both small ($k \approx 1$) and very large ($k \rightarrow \infty$) values of $k$ appear to respect the conformal limit, intermediate values—most notably $k \approx 3.0$—show the speed of sound rising above the conformal limit ($v_s^2 > 1/3$) for temperatures $T/T_c > 1.5$. This suggests that the RGPNJL framework introduces a non-trivial competition between the chiral sector and the Polyakov loop dynamics. This parametric sensitivity indicates that simply increasing $k$ does not uniformly suppress the speed of sound towards the conformal limit in the presence of Polyakov loops, pointing to a specific parametric region that warrants further investigation to reconcile the RG-consistent requirements with the underlying thermodynamics of the model.

The RGPNJL model (Fig.~\ref{PNJL4}d) demonstrates a more intricate, non-monotonic dependence on $k$. While both small ($k \approx 1$) and very large ($k \rightarrow \infty$) values of $k$ appear to respect the conformal limit, intermediate values—most notably $k \approx 3.0$—show the speed of sound rising above the conformal limit ($v_s^2 > 1/3$) for temperatures $T/T_c > 1.5$. This suggests that the RGPNJL framework introduces a non-trivial competition between the chiral sector and the Polyakov loop dynamics. This parametric sensitivity indicates that simply increasing $k$ does not uniformly suppress the speed of sound towards the conformal limit in the presence of Polyakov loops, pointing to a specific parametric region that warrants further investigation to reconcile the RG-consistent requirements with the underlying thermodynamics of the model.

\begin{figure}[!htbp]
	\centering
	\subfigure[]{
		\label{tensorSp1}
		\includegraphics[scale=0.45]{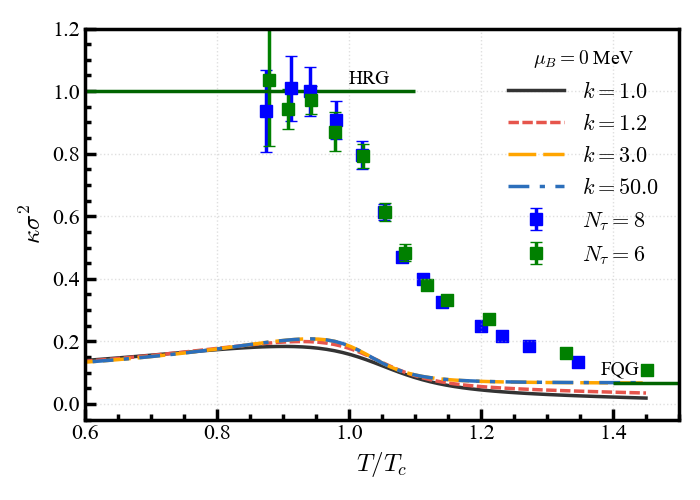}}
	\subfigure[]{
		\label{tensorSp2}
		\includegraphics[scale=0.45]{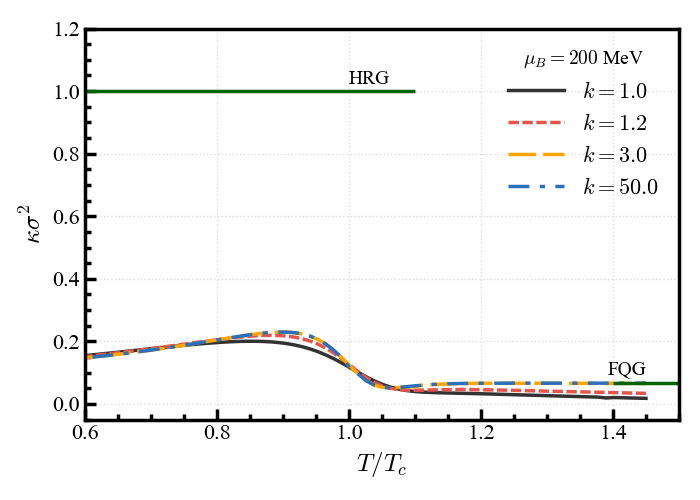}}
	\subfigure[]{
		\label{tensorSp1}
		\includegraphics[scale=0.45]{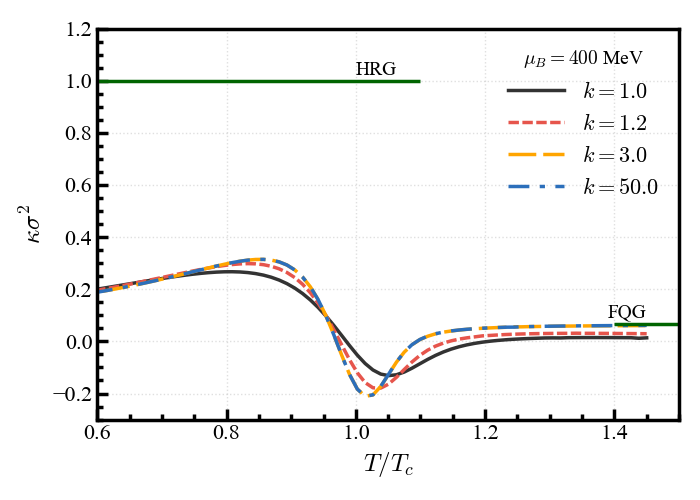}}
	\subfigure[]{
		\label{tensorSp2}
		\includegraphics[scale=0.45]{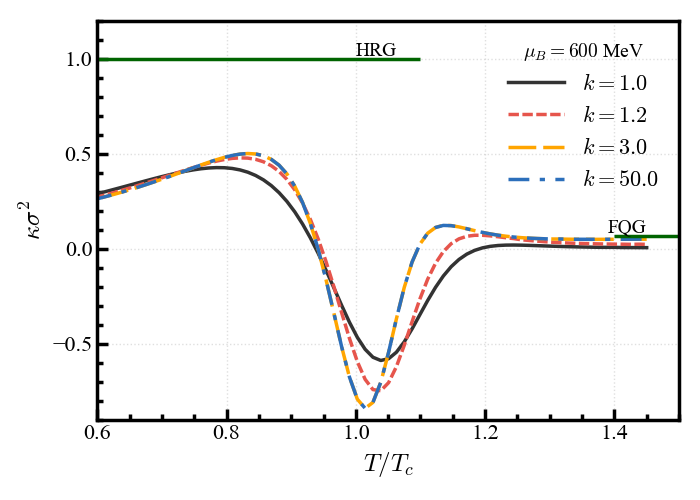}}
	\caption{Net-baryon number kurtosis $\kappa\sigma^{2}$ as a function of temperature $T/T_c$ for various truncation factors $k$ (where $\Lambda_T = k\Lambda_0$) at fixed baryon chemical potentials $\mu_B = 0, 200, 400,$ and $600$~MeV in the RGNJL model. The results for $\mu_B=0$ are compared with lattice QCD data from \cite{Bazavov:2017dus}. The horizontal lines represent the hadron resonance gas (HRG) limit ($\kappa\sigma^{2}=1$) and the ideal free quark gas (FQG) limit ($\kappa\sigma^{2}=0.068$).}
	\label{NJLks}
\end{figure}

\par Complementary to the above thermodynamic analysis, the nature of the chiral phase transition can be further elucidated by examining higher-order fluctuations of the net-baryon number. These fluctuations are defined through generalized susceptibilities, $\chi_{n}^{B}$, which are derived from the thermodynamic potential $\Omega$ (where $P = -\Omega$):
\begin{equation}
	\chi_{n}^{B} = \frac{\partial^{n}[P/T^{4}]}{\partial[\mu_{B}/T]^{n}}.
\end{equation}
The corresponding cumulants of the baryon number distributions, $C_{n}^{B}$, are then expressed as:
\begin{equation}
	C_{n}^{B} = V T^{3} \chi_{n}^{B}.
\end{equation}
By defining the variance $\sigma^{2} = C_{2}^{B}$ and the kurtosis $\kappa = C_{4}^{B} / (\sigma^{2})^{2}$, one can construct the ratio $\kappa \sigma^{2}$, which serves as a vital bridge between theoretical framework predictions and experimental observables:
\begin{equation}
	\kappa \sigma^{2} = \frac{C_{4}^{B}}{C_{2}^{B}} = \frac{\chi_{4}^{B}}{\chi_{2}^{B}}.
\end{equation}
This ratio, $\kappa\sigma^{2}$, is a highly sensitive indicator of the criticality of the QCD phase transition~\cite{Fu:2021oaw,Fu:2021wyc,Lu:2025qyf}. Utilizing these ratios allows for a rigorous quantitative comparison between the RGNJL/RGPNJL frameworks and lattice QCD data, providing an essential benchmark to determine whether the RG-consistent modifications accurately capture the fluctuation properties of the hot, dense medium.

\par
\begin{figure}[!htbp]
	\centering
	\subfigure[]{
		\label{tensorSp1}
		\includegraphics[scale=0.45]{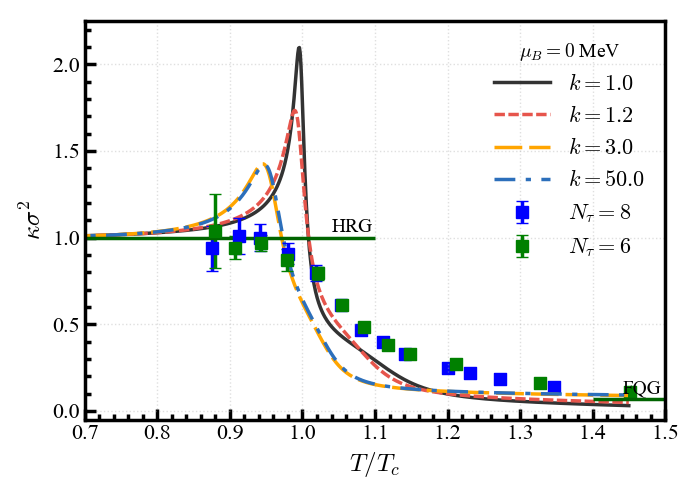}}
	\subfigure[]{
		\label{tensorSp2}
		\includegraphics[scale=0.45]{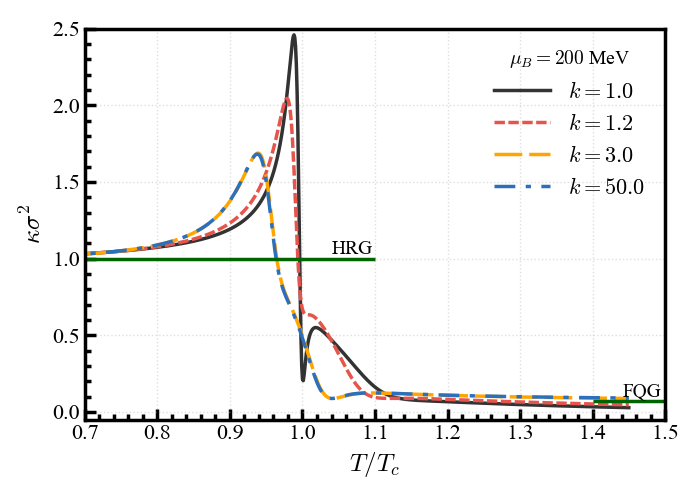}}
	\subfigure[]{
		\label{tensorSp1}
		\includegraphics[scale=0.45]{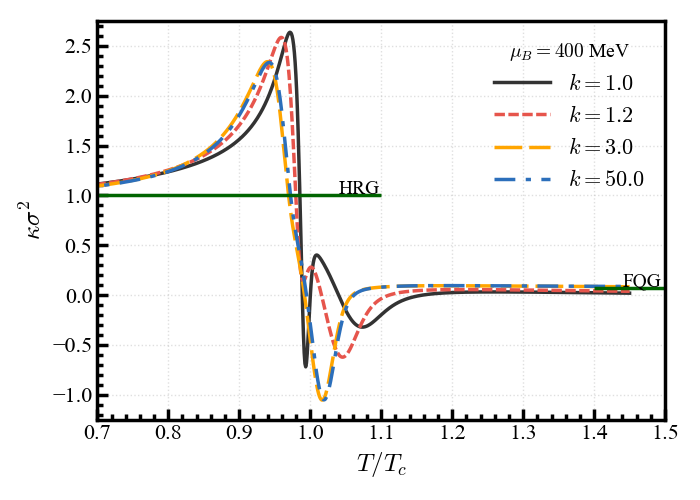}}
	\subfigure[]{
		\label{tensorSp2}
		\includegraphics[scale=0.45]{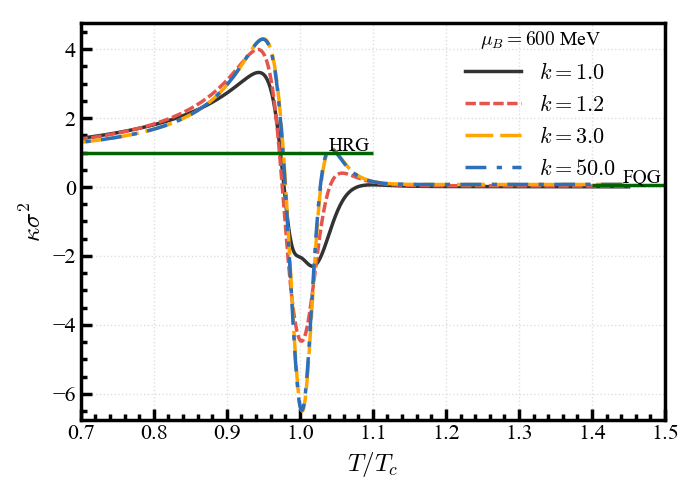}}
	\caption{Net-baryon number kurtosis $\kappa\sigma^{2}$ as a function of temperature $T/T_c$ for various truncation factors $k$ (where $\Lambda_T = k\Lambda_0$) at fixed baryon chemical potentials $\mu_B = 0, 200, 400,$ and $600$~MeV in the RGPNJL model. The results for $\mu_B=0$ are compared with lattice QCD data from \cite{Bazavov:2017dus}. The horizontal lines represent the hadron resonance gas (HRG) limit ($\kappa\sigma^{2}=1$) and the ideal free quark gas (FQG) limit ($\kappa\sigma^{2}=0.068$).}
	\label{PNJLks}
\end{figure}

 As illustrated in Fig.~\ref{NJLks} and Fig.~\ref{PNJLks}, the behavior of the net-baryon number kurtosis $\kappa\sigma^2$ exhibits distinct characteristics depending on the model and the chemical potential. At vanishing chemical potential ($\mu_B = 0$), the kurtosis in both models remains relatively stable in the low-temperature regime, consistent with the Hadron Resonance Gas (HRG) limit where $\kappa\sigma^2 \approx 1$. As temperature increases, the RGNJL model (Fig. 6a) shows minimal sensitivity to the RG parameter $k$ and significantly underestimates the lattice simulation data in the transition region. In contrast, the RGPNJL model (Fig. 7a) provides a much better qualitative match to the lattice results. In particular, as the RG consistency condition is approached ($k \rightarrow \infty$), the kurtosis in the RGPNJL model exhibits a more rapid decrease for $T/T_c \ge 1.1$, aligning closely with the lattice trend as it approaches the ideal free quark gas (FQG) limit of approximately $0.068$.

\par However, as the chemical potential $\mu_B$ increases, the kurtosis develops a severe and non-monotonic dependence on both temperature and the parameter $k$. In the range $0.8 \le T/T_c \le 1.2$, the fluctuations become increasingly volatile: a larger $k$ value leads to a higher maximum and a deeper minimum in the kurtosis. Interestingly, increasing $\mu_B$ causes the peak position of the kurtosis to shift toward lower temperatures, a feature observable in both models but significantly magnified in the RGPNJL framework (Fig. 7c and 7d). While the RGNJL model (Fig. 6) maintains relatively smooth oscillations even at high $\mu_B$, the RGPNJL model displays increasingly sharp, "spike-like" structures. For $\mu_B = 600$ MeV, the RGPNJL model with $k=50.0$ displays a sharp oscillation from a positive peak near $\kappa\sigma^2 \approx 4$ to a negative dip below $-6$. This dramatic intensification of the extrema near $T/T_c \approx 1.0$ suggests that the coupling between the Polyakov loop and the RG-consistent chiral dynamics heightens the model's sensitivity to the critical point's proximity at high baryon density. These shifts indicate that while RG consistency improves high-temperature thermodynamics, it simultaneously amplifies the signatures of critical fluctuations in the dense medium.

\section{Conclusion and Discussion}
\label{sec:5}

\par In this study, we have systematically investigated the impacts of renormalization group (RG) consistency on the thermodynamic properties and chiral phase transitions within the RGNJL and RGPNJL frameworks. Our results demonstrate that the implementation of a separate thermal cutoff $\Lambda_T = k\Lambda_0$ significantly modifies the behavior of the constituent quark mass, critical temperature $T_c$, and higher-order baryon number fluctuations. By increasing the truncation factor $k$, both models successfully converge toward the expected Stefan-Boltzmann limits at high temperatures, addressing a long-standing limitation of fixed-cutoff effective theories. This convergence confirms that the RG-consistency framework provides a robust method for extending the applicability of non-renormalizable models into the high-energy regime.

\par A comparative analysis between the RGNJL and RGPNJL models reveals both distinct advantages and inherent limitations. The primary advantage of the RG-consistent approach is most evident in the RGNJL model's speed of sound, where the enforcement of $k \rightarrow \infty$ effectively resolves the unphysical causality violation ($v_s^2 > 1$) present in the standard NJL baseline. Furthermore, the RGPNJL model exhibits superior agreement with lattice QCD data for baryon number kurtosis at vanishing chemical potential, particularly in capturing the sharp transition dynamics near $T_c$. However, a significant disadvantage persists in both models: for large $k$, the constituent quark mass can drop below the current quark mass ($M < m$) in the restored phase. This unphysical artifact suggests that while the RG method improves thermal contributions, the current regularization of the vacuum sector and the transition between vacuum and thermal modes require further refinement to maintain the fundamental bounds of the dynamical mass.

\par Furthermore, the behavior of the squared speed of sound and kurtosis extrema in the RGPNJL model points to a non-trivial competition between chiral dynamics and the Polyakov loop potential. We observed that intermediate values of $k$ (e.g., $k \approx 3.0$) can lead to $v_s^2$ exceeding the conformal limit in the RGPNJL framework, a feature not seen in the RGNJL case. This sensitivity suggests that the inclusion of confinement effects through the Polyakov loop introduces additional constraints on the RG parameter space. In the dense medium ($\mu_B > 400$ MeV), the RG-consistent modifications amplify the magnitude of kurtosis oscillations, particularly in the RGPNJL model. While this intensification provides a more sensitive probe for searching for the QCD critical point, it also highlights the model's high dependency on the specific parametric choice of the truncation factor in regimes where non-perturbative effects are dominant.

\par In conclusion, the RG method serves as an essential improvement for NJL-type models by ensuring that high-momentum thermal modes are adequately accounted for, thereby aligning effective theory predictions with the fundamental requirements of QCD thermodynamics. The RGPNJL model, in particular, stands out as a more reliable tool for phenomenological comparisons with experimental fluctuation data from the STAR collaboration, provided that $k$ is chosen to satisfy the conformal limit. Future work should focus on developing a more sophisticated interpolation between the vacuum and thermal regulators to eliminate unphysical mass artifacts and on extending this RG-consistent framework to the $(2+1)$-flavor case to incorporate the effects of strangeness on the QCD phase diagram.

\acknowledgements
 F.L. acknowledges support from the National Natural Science Foundation of China (Grant No. 12547138) and the Anhui University of Science and Technology (Grant No. 2025yjrc0143). X.W. was supported by the Anhui University of Science and Technology under Grant No. YJ20240001.

\bibliographystyle{unsrt}
\bibliography{references}

\end{document}